\documentclass[prd,showpacs,twocolumn,superscriptaddress]{revtex4}
\usepackage{graphicx}

\def\be{\begin{equation}}
\def\ee{\end{equation}}
\def\bea{\begin{eqnarray}}
\def\eea{\end{eqnarray}}

\def\Rc{{\cal{R_{\rm{c}}}}}
\def\Dc{{\Delta_{\rm{th}}}}

\def\lsim{\mathrel{\rlap{\lower4pt\hbox{\hskip1pt$\sim$}}
    \raise1pt\hbox{$<$}}}
\def\gsim{\mathrel{\rlap{\lower4pt\hbox{\hskip1pt$\sim$}}
    \raise1pt\hbox{$>$}}}
\def\esim{\mathrel{\rlap{\raise2pt\hbox{\hskip0pt$\sim$}}
    \lower1pt\hbox{$-$}}}

\begin{document}
\title{Primordial black hole constraints on non-gaussian inflation models}
\author{Pedro P. Avelino}
\affiliation{Centro de F\'{\i}sica do Porto e Departamento de F\'{\i}sica da 
Faculdade de Ci\^encias da Universidade do Porto, Rua do Campo Alegre 687, 
4169-007, Porto, Portugal}
\date{\today}

\begin{abstract}

We determine the abundance of primordial black holes (PBHs) formed in the 
context of non-gaussian models with primordial density perturbations. We 
consider models with a renormalized $\chi^2$ probability 
distribution function parametrized by the number, $\nu$, of degrees of 
freedom. We show that if $\nu$ is not too large then the PBH 
abundance will be altered by several orders of magnitude with respect 
to the standard gaussian result obtained in the  
$\nu \to \infty$ limit. We also study the dependence of the 
spectral index constraints on the 
nature of the cosmological perturbations for a power-law primordial 
power spectrum. 

\end{abstract}

\maketitle
%%%%%%%%%%%%%%%%%%%%%%%%%%%%%%%%%%%%%%%%%%%%%%%%%%%%%%%%%%%%%%%%%%%%%%

\section{Introduction}

Contrary to early beliefs some level of 
non-gaussianity is expected to be present in all 
inflationary models. Furthermore, 
the non-gaussian contribution will in general be scale-dependent (see 
for example \cite{BKMR} and references therein). Consequently, 
a strong non-gaussian component on small-scales may turn out to 
be consistent with current observations which only allow for small deviations 
from gaussianity on large cosmological scales \cite{Ketal}. 
Topological defect models also produce a scale-dependent non-gaussianity 
in a natural way \cite{ASWA}. However, in these models the fluctuations are 
not primordial but generated actively on increasing larger scales 
by defect network evolution.

Black holes of various masses are expected to form in the 
early universe as a result of the collapse of large density 
perturbations \cite{H1,ZN}.
The abundance of PBHs provides useful constraints on the 
primordial power spectrum over a wide range of small scales 
(see for example \cite{C,CGL,GL,GLMS,C1} and references there in). 
Although, the usual 
calculation assumes gaussian 
perturbations (see however ref. \cite{BP} for a discussion of the role 
of non-gaussian fluctuations in PBH formation), standard constraints will 
need to be modified when considering non-gaussian models.

The cosmological constraints on the abundance of primordial black holes 
come either from their contribution to the matter density at the present 
time or from the cosmological implications which result from their 
Hawking evaporation \cite{H2}. PBHs with mass $M \lsim 10^{15}$ will 
have evaporated by the present epoch but if they have a mass $M \gsim 10^9$ 
they will still be present at the nucleossynthesis epoch. Assuming a standard 
cosmological scenario, the fraction of the universe energy density in 
PHBs with mass greater than $M$  at the time they form is 
constrained by observations to be $\lsim 10^{-20}$ on any interesting mass 
scale with more accurate constraints depending on 
their mass (see for example ref. \cite{C1}). 

We will generalize to the non-gaussian case the standard 
calculation of primordial black hole abundances. This uses the 
Press-Schechter approach \cite{PS} in order to estimate the fraction of 
the energy 
density in PBHs. An alternative calculation using peaks 
theory 
was performed in \cite{GLMS} using the criterion for black hole formation 
derived 
by Shibata and Sasaki \cite{SS} which 
refers to the the metric perturbation rather than the density field. 
However, given the large uncertainties involved the results obtained by the 
two methods were shown to be consistent.

\section{The power spectrum}

On comoving hypersurfaces there is a simple relation between the density
perturbation, $\Delta$, and the curvature perturbation, $\Rc$ 
(e.g. \cite{LL}) 
\be
\Delta(t,k)=\frac{2(1+w)}{5+3w}\left(\frac{kc}{aH}\right)^2{\Rc}(k)\,.
\ee
Here $w=p/\rho$ is the equation of state, $H={\dot a}/a$ is the 
Hubble parameter, $k$ is a comoving wavenumber and $c$ is the speed of 
light. The power spectra 
are related by
\be
{\cal P}_{\Delta}(k)=\frac{4(1+w)^2}{(5+3w)^2}\left(\frac{kc}{aH}\right)^4
{\cal P}_{{\Rc}}(k)\,,
\ee
and consequently at horizon crossing 
\be
\label{horcross}
{\cal P}_{\Delta}(k) 
= \frac{4(1+w)^2}{(5+3w)^2} {\cal P}_{{\Rc}}(k) \,.
\ee
If we assume a power-law primordial power spectrum with 
${\cal P}_{\Rc}(k) = A_{\Rc} (k/k_{0})^{n-1}$ and choose a gaussian window 
function, 
\be 
W(kR)=\exp{\left(-\frac{ k^2 R^2}{2}\right)}\,, 
\ee
then the variance of the primordial density field smoothed on a scale $R$ 
at horizon crossing,
\be
\sigma_{\Delta}^2(R)=\int_{0}^{\infty} W^{2}(kR){\cal P}_{\Delta}(k) 
\frac{{\rm d}k}{k} \,, 
\ee
is given by
\be
\sigma_{\Delta}^2(M)=\frac{2(1+w)^2}{(5+3w)^2}
\frac{A_{{\Rc}} \Gamma[(n-1)/2]}{(k_{0} R)^{n-1}}\,,
\ee
for $n>1$. We note that for $n \le 1$ the variance of a density 
field with a power-law primordial power-spectrum diverges due to 
contributions near $k=0$. 
However, a small $k$ (large wavelength) cut-off to the power spectrum is 
expected on physical grounds.
Spergel et al. \cite{Setal} have shown, using the WMAPext+2dFGRS dataset,
that $A_{\Rc}=(0.8 \pm 0.1) \times 2.95 \times 10^{-9}$ for
$k_{0}=0.05 {\rm Mpc}^{-1}$. We shall take $w=1/3$ appropriate for the 
radiation dominated era.

\section{The PBH abundance}

We shall use the Press--Schechter approximation (\cite{PS}) to 
compute the fraction of the energy of the Universe, $f(> M)$, associated 
with PBHs with masses larger than $M$ at the time they form. 
The Press--Schechter approximation was originally
proposed in the context of initial gaussian density perturbations and was much
later generalized to accommodate non-gaussian initial conditions 
\cite{COS}.

The mass fraction is assumed to be
proportional to the fraction of space in which the linear density
contrast, smoothed on the scale $M$, exceeds a given threshold $\Dc$: 
\be 
\label{mf}
f(> M)= A_f \int_\Dc^\infty  P_R(\Delta) d
\Delta \,.  
\ee 
Here $P_R(\Delta)$ is the probability
distribution function (PDF) of the linear density field $\Delta$ smoothed on 
a scale $R$ and $A_f$ is assumed to be a constant which can be calculated by 
requiring that $f(>0)=1$, thus taking into account the accretion of material 
initially present in underdense regions (note that $A_f=2$ in the case of 
gaussian initial conditions). 

This generalization has successfully reproduced the results obtained from 
$N$-body simulations with non-gaussian initial conditions 
\cite{RB} but it has been shown in ref. \cite{AV} that 
it does not adequately solve the cloud-in-cloud problem (see also ref. 
\cite{IN}). Although this means that in many models $A_f$ is not expected 
to be exactly constant, it will in general be of order unity. Consequently 
this small uncertainty will have a negligible impact on our final 
results and we shall not consider it further in the following analysis 
(we will use $A_f=1$ throughout the paper).

The precise value of the threshold value $\Dc$ relevant for the calculation of 
the PHB abundance is uncertain but we shall use 
the standard value, $\Dc=1/3$, for a radiation 
dominated universe \cite{C}. Although there should also be a finite upper limit to 
the integration in Eq.~(\ref{mf}) in practice the upper cut-off is unimportant 
since $P_R(\Delta)$ is usually a rapidly decreasing function of $\Delta$.

In order to implement our calculation one needs to relate the mass 
$M$ to the comoving smoothing scale $R$. An overdense region in order 
to collapse must be larger than the Jeans length at maximum expansion which, 
in the radiation dominated era, is $3^{-1/2}$ times the horizon size. 
On the other hand, 
it cannot be larger than the horizon size or otherwise it would form a
closed universe separate from our own \cite{CH}. For simplicity, here we 
will assume 
that PBHs will form with roughly the horizon mass. 
This assumption is only approximately true (in general the mass of primordial 
black holes is expected to depend on the amplitude, size and shape of the 
perturbations). It will nevertheless have a small impact on our 
main results. Hence, 
when the comoving scale $R=c(Ha)^{-1}$ enters the horizon the horizon mass 
is 
\be
M_H= \frac{4\pi}{3}\rho (c/H)^{-3} \,,
\ee
and primordial black holes with a mass $M=M_H$ may form at that time. In 
the radiation dominated era $\rho \propto g_\star^{-1/3} a^{-4}$ where 
$g_\star$ is the number of relativistic degrees of freedom 
(we approximate the temperature and entropy degrees of freedom as equal).
This implies that 
\be
M_H= \frac{c}{2G}\left(\Omega_{\gamma,0}\right)^{1/2} 
H_0 R^2 \left(\frac{2}{g_{\star}} \right)^{1/6}\,.
\ee
Taking into account that $\Omega_{\gamma,0} h^2=2.5 \times 10^{-5}$ it is 
straightforward to show that  
\be
\label{rm}
\frac{R}{1 \, {\rm {Mpc}}}=5 \times 10^{-24} 
\left(\frac{M}{1 \, {\rm g}}\right)^{1/2} 
\left(\frac{g_{\star}}{2} \right)^{1/12} \,.
\ee
The number of relativistic degrees of freedom, 
$g_{\star}$, in the early universe is expected to be of order $100$. 
However, the smoothing scale $R$ is only weakly dependent on $g_{\star}$ 
(note that Eq.~(\ref{rm}) corrects a minor error in ref. \cite{GLMS}).

\section{The $\chi^2$ model}

We consider an initial density field with a chi-squared 
probability distribution function (PDF) with $\nu$ degrees of freedom, 
the PDF having been shifted so that its mean is zero (such a PDF 
becomes gaussian when $\nu \to \infty$). Hence,
\be
f(> M) = Q(\nu/2,\Delta_{th}/\sigma_\Delta \times {\sqrt {\nu/2}}+\nu/2) \,,  
\ee 
where $Q(a,x)$ with $a>0$ is the incomplete gamma function defined by
\be
Q(a,x) \equiv \frac{\Gamma(a,x)}{\Gamma(a)} \,.
\ee
Here
\be
\Gamma(a,x) \equiv \int_x^\infty e^{-t} t^{a-1} dt 
\ee
and $\Gamma(a)=\Gamma(a,0)$. If $\nu \to \infty$ then  
\be
f(> M) \to \frac{1}{2} {\rm erfc}\left( \frac{\Delta_{th}}{{\sqrt 2} \sigma_\Delta}\right) \,.  
\ee 
We further
assume that the shape of the PDF is scale independent, that is 
$P_R(\Delta) \equiv P (R,\Delta/\sigma_\Delta(R))/\sigma_\Delta(R)$ is always 
the same function (or equivalently $\nu$ does not depend on $R$). 
Although this is not expected to be precisely true in 
realistic models, it may be a good approximation over the range of scales 
relevant for the determination of PBH abundances.

\begin{figure}[t!]
\includegraphics[width=8.5cm]{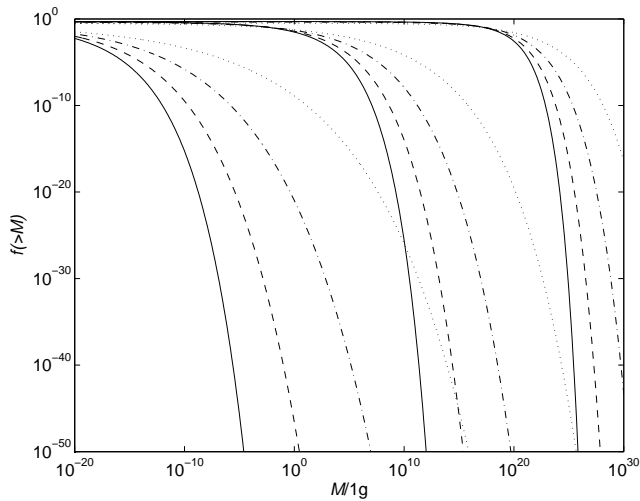}\\
\caption[fig1]{\label{fig1} PBH abundance as a function of black hole mass for 
power-law power spectra with $n=1.2$, $n=1.3$ and $n=1.5$ 
(sets of curves from left to right respectively) in a $\chi^2$ model 
with $n=\infty$, $n=100$, $n=10$ and $n=1$ degrees of freedom 
(solid, dashed, dot-dashed and dotted lines respectively).}
\end{figure}

\begin{figure}[t!]
\includegraphics[width=8.5cm]{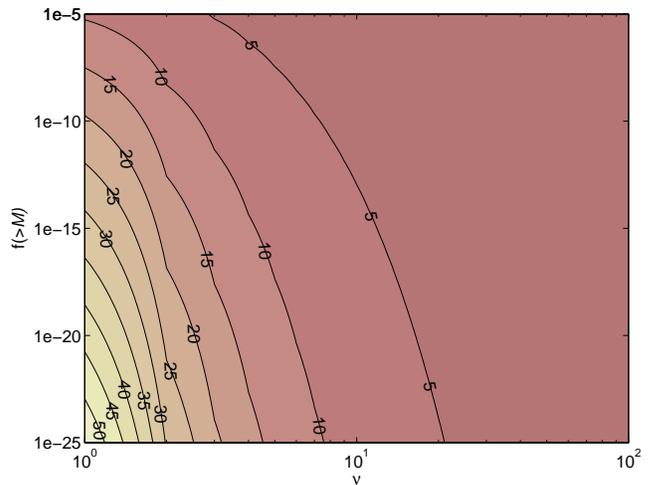}\\
\caption[fig2]{\label{fig2} The value of $\alpha$ (defined as the ratio 
between the amplitudes of the perturbations in a standard gaussian model 
and in a $\chi^2$ model with identical mass functions, $f(> M)$, at a 
given mass, $M$) as a function of $\nu$ and  $f(> M)$. Note that as 
expected $\alpha$ decreases with increasing $\nu$ 
(approaching $1$ when $\nu \to \infty$). Also, for a fixed $\nu$ a 
smaller $f(> M)$ implies a larger $\alpha$.
}
\end{figure}

\section{Results and discussion}

In Fig.~\ref{fig1} we plot the PBH abundance as a function of black hole mass 
assuming a power-law spectra with $n=1.2$, $n=1.3$ and $n=1.5$ 
(sets of curves from left to right respectively). The gaussian model is 
represented by a solid 
line and the dashed, dot-dashed and dotted lines represent $\chi^2$ 
models with $n=100$, $n=10$ and $n=1$ degrees of freedom respectively. 
We clearly see that even a small amount of non-gaussianity may have a large 
impact on the estimated PBH abundance. This shows that an improper use 
of the standard gaussian assumption may introduce errors on the estimated 
PBH abundance of many orders of magnitude. 

It is useful to determine the 
impact of non-gaussianity on the spectral index constraint, $n$. We clearly 
see from Fig.~\ref{fig1} that for $\chi^2$ models with a finite numbers of 
degrees of freedom the constraints are strengthened since for the same 
value of the spectral index the PHB abundance is significantly increased. 
Still, this will at most shift the spectral index constraint down by about 
$0.1$ (in the most extreme case with $\nu=1$).

It is also interesting to ask how much larger the amplitude of the 
perturbations has to be in a standard gaussian model relative to a 
$\chi^2$ model in order for the mass functions $f(> M)$ at a given mass 
to be the same.
An answer to this question is easily obtained by solving the equations
\bea
\frac{1}{2} {\rm erfc}\left(x\right) &  =& f(> M) \,, \nonumber \\
     Q(\nu/2, x \alpha {\sqrt {\nu}}+\nu/2)   & = & f(> M) \,.
\eea
with respect to $\alpha$ for a given $\nu$ and $f(> M)$. The results are 
shown in Fig.~\ref{fig2}. As expected we clearly see that if $\nu$ is 
increased then $\alpha$ decreases (approaching $1$ when $\nu \to \infty$).
Also, for a fixed $\nu$ a smaller $f(> M)$ implies a larger $\alpha$ 
since the differences between gaussian and $\chi^2$ probability 
distribution functions (for $\nu \neq \infty$) become 
more dramatic as we move deeper into the distribution tail.

In this study we do not compute the PBH for any specific model of inflation. 
Still, we expect the $\chi^2$ class of models we investigate in this paper 
to be representative of a large family of non-gaussian models which may be 
realized in the context of specific inflationary scenarios. The simple 
assumptions made in 
this paper also have the advantage of allowing for a 
complete separation between the effect of the power spectrum 
(which controls the amplitude of 
density perturbations on the scales relevant for the calculation of PBHs 
abundances) from the distribution of the phases 
of the various Fourier modes. Although, it is possible 
to consider stronger deviations from a gaussian random distribution which 
would modify the constraints even further, it might be difficult to construct 
realistic models where such deviations are incorporated in a natural way.

\section*{ACKNOWLEDGMENTS}

P.P.A. was partially supported by Funda{\c c}\~ao para a Ci\^encia e a
Tecnologia (Portugal) under contract POCTI/FP/FNU/50161/2003.

\end{document}